\documentclass[a4paper]{article}

\usepackage{INTERSPEECH2022}
\usepackage{todonotes}
\usepackage{url}
\usepackage{tabularx}
\usepackage{xspace} 
\usepackage{graphicx}
\usepackage{multirow}
\usepackage{hyperref}
\usepackage{xcolor}
\usepackage{algorithm}
\usepackage{algpseudocodex}
\newcommand{\MI}{MI\xspace}

\usepackage{pdfpages}
\DeclareRobustCommand\encircle[1]{\tikz[baseline=(char.base)]{\node[shape=circle,fill,inner sep=1pt] (char) {\textcolor{white}{#1}}}}

\usepackage{pgf}

\DeclareMathOperator*{\argmin}{argmin}

\title{Introducing Model Inversion Attacks on Automatic Speaker Recognition}
\name{Karla Pizzi$^{*,1,2}$\thanks{$^{*}$Authors contributed equally.}, Franziska Boenisch$^{*,1}$, Ugur Sahin$^{*,1,2}$, Konstantin Böttinger$^1$}
\address{
  $^1$Fraunhofer AISEC, Germany;
  $^2$Technical University Munich, Germany}
\email{[firstname.lastname]@aisec.fraunhofer.de}

\begin{document}

\maketitle
\begin{abstract}
  Model inversion (\MI) attacks allow to reconstruct average per-class representations of a machine learning (ML) model's training data.
  It has been shown that in scenarios where each class corresponds to a different individual, such as face classifiers, this represents a severe privacy risk. 
  In this work, we explore a new application for \MI: the extraction of speakers' voices from a speaker recognition system.
  We present an approach to (1)~reconstruct audio samples from a trained ML model and (2)~extract intermediate voice feature representations which provide valuable insights into the speakers' biometrics.
  
  Therefore, we propose an extension of \MI attacks which we call \emph{sliding model inversion}.
  Our sliding \MI extends standard \MI by iteratively inverting overlapping chunks of the audio samples and thereby leveraging the sequential properties of audio data for enhanced inversion performance.
  We show that one can use the inverted audio data to generate spoofed audio samples to impersonate a speaker, and execute voice-protected commands for highly secured systems on their behalf.
  To the best of our knowledge, our work is the first one extending \MI attacks to audio data, and our results highlight the security risks resulting from the extraction of the biometric data in that setup.
\end{abstract}

\section{Introduction}


Privacy analysis of audio data has shown that speech parameters, such as accent, rhythm, or acoustic properties of speech inherently carry biometric information about the speakers, such as their age, gender, physical health, and geographical origin~\cite{kroger2019privacy}.
Therefore, it is important for machine learning (ML) models in speaker recognition not to leak information about their training data. 
However, recent research~\cite{fredrikson2015model,shokri2017membership, ganju2018property} suggests that ML models are, in general, vulnerable to privacy attacks. 
One particular attack is \emph{model inversion}~(\MI)~\cite{fredrikson2015model} which allows an attacker to retrieve abstract representations for individual classes of the target model's training data.
With speaker recognition systems treating each individual as their own class, \MI attacks have the potential to cause severe privacy breaches~\cite{fredrikson2015model}. 
So far, the feasibility of \MI attacks on speaker recognition systems and audio data has never been tested, thus, the question if information on the speakers can be maliciously retrieved remained open.

We are the first to show how to adapt and apply \MI attacks for audio data.
We do so by targeting SincNet, a state-of-the-art neural network (NN)-based speaker recognition model \cite{ravanelli2018speaker,ravanelli2021speech}. 
We show that \MI attacks are able to infer both entire audio samples and d-vectors as intermediate representations of the speakers' voice characteristics from the trained target model. 
Further, we propose the \emph{sliding model inversion}, a novel form of the standard \MI attack that leverages sequential processing properties of the audio data to improve inversion success. 
While with standard \MI, the target model successfully identifies up to $54$\% of the inverted audio samples as their correct speaker class, with our novel attack, we achieve to up to $90\%$ accuracy.
Also, our sliding \MI manages to decrease the distance between original and inverted sample in the d-vector representation, hence, yielding higher fidelity inversions. 
For directly inverting d-vectors, our experiments show that even standard \MI achieves $100\%$ identification success.
These results highlight the vulnerability of speaker recognition models to privacy attacks. 
As a proof-of-concept to showcase that our \MI can be exploited as a departure point for further attacks against speaker recognition, we, furthermore, explore using inverted audio samples as inputs for deepfake generation.
Such deepfakes could be used to fool voice identification with arbitrary speech samples or to execute any speech command on behalf of the speakers under attack.
While our generated deepfakes do not perfectly fool a human listener, as an informal evaluation conducted by the authors shows, they illustrate that privacy attacks can not only be used to disclose sensitive information about the individuals the model was trained on;
additionally, they can severely threaten the security of systems relying on voice biometrics. 
Our contributions can be summarized as follows:
\begin{itemize}
    \item We successfully apply \MI attacks on speaker recognition models to invert entire audio samples and d-vectors and experimentally evaluate what kind of random initialization works best as an input for \MI attacks on audio data.
    \item We introduce a novel \emph{sliding \MI} which exploits properties of sequential and chunk-wise audio processing.
    \item We show the feasibility of generating deepfakes based on the inferred audio samples. 
\end{itemize}

\section{Background and Related Work}

The following section provides background information on speaker recognition systems and attacks against their privacy.

\begin{figure}[t]
    \centering
    \includegraphics[trim=0 3.5cm 0 0,clip,width=\linewidth, page=1]{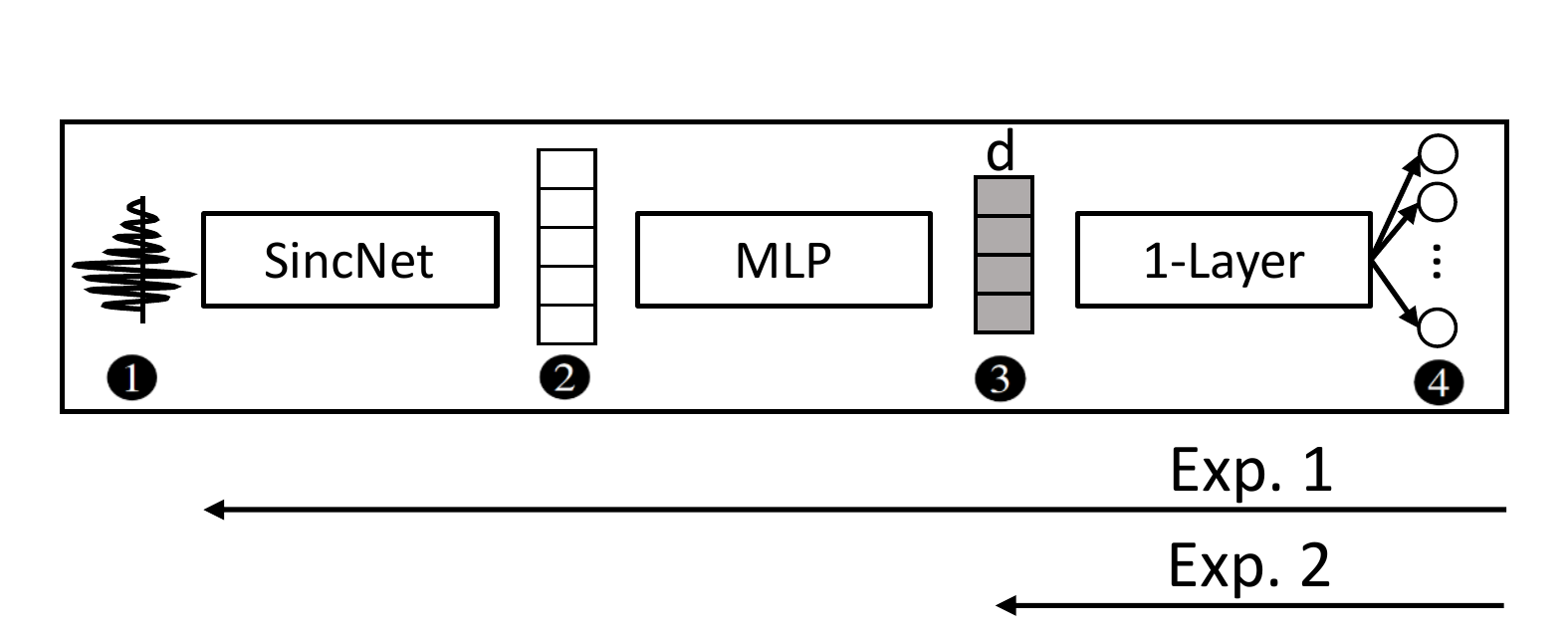}
    \caption{\textbf{Speaker Recognition Model.} The speaker recognition model and its three submodels: 
    SincNet obtains \encircle{1} raw audio input and generates \encircle{2} features. 
    These features are input to an MLP which generates the \encircle{3} d-vectors. 
    A single layer performs \encircle{4} classification on them.
    In experiment~1, we invert full audio samples, while in experiment~2, we invert the d-vectors.}
    \label{fig:experiments}
\end{figure}

\paragraph*{Speaker recognition.}
\label{ssec:speakerrec}
In this paper, we use a SincNet-based~\cite{ravanelli2018speaker} text-independent speaker recognition system.
This system uses NNs to extract voice features into so-called d-vectors and adds a classification layer on top of these.
The input to the system consists of raw audio waves and the outputs is a per-class probability score over all possible classes (i.e., speakers).
Overall, the system is composed of three submodels (see Figure~\ref{fig:experiments}) 1) SincNet, a convolutional NN that resembles a band-pass filter; 2) a multi-layer perceptron (MLP) calculating the d-vectors \cite{variani2014deep}; and 3) a fully connected layer to calculate the probabilities per speaker.
It achieves a reported classification error\footnote{See \url{https://pythonlang.dev/repo/mravanelli-sincnet/}.} of $5.772 \cdot 10^{-3}$ on the TIMIT~\cite{garofolo1993timit} test data set (measured at sentence level).
In its current version, SincNet does not provide any dedicated privacy-preserving mechanisms.

\paragraph*{Privacy in speaker recognition.}
\label{ssec:speakerprivacy}
The ISO/IEC norm 24745:2011 proposes three general requirements to ensure individual privacy: irreversibility, renewability, and unlinkability of the protected data \cite{nautsch2019preserving}.
To reach this goal in speaker recoginition systems, several solutions have been proposed and discussed \cite{nautsch2019preserving}. 
However, aiming for any \emph{anonymization} to protect the privacy of speakers in a speaker recognition model goes directly against the purpose of the system, which is to identify speakers based on their individual characteristics.

\paragraph*{\MI attacks.}
\label{ssec:modelinv}

In \MI attacks~\cite{fredrikson2015model}, the attacker exploits an ML model's prediction confidence for inverting individual training classes (see Algorithm~\ref{algo:standardMI}).
More formally speaking, a \MI attack can be expressed as follows:
let $f$ be the target model under attack.
It is trained to map from an $n$-dimensional input data point $x$ to an $m$-dimensional vector $p$ indicating the probability per class, such that
$f: x \mapsto p$, with $\mathbb{R}^n \rightarrow [0,1]^m$.
To invert the model, we define an objective function in order to use gradient descent.
This function is called cost function $c(x)$ and basically defines how close we are to the information we would like to reconstruct. 
We set $c(x) = 1 - p_{t}$, where $t$ denotes the target class we would like to gain information about. 
Starting from a randomly initialized input sample $x_0$, we calculate its cost $c(x_0)$.
With this at hand, we apply the gradient descent algorithm for $\alpha$ iterations with a learning rate $\lambda$ to alter the original input.
The aim is to minimize the costs for a specific class~\cite{fredrikson2015model}, such that the resulting data sample is a representation of that class.
\begin{algorithm}[tb]
\caption{\emph{Standard} Model Inversion Attack~\cite{fredrikson2015model}}\label{alg:cap_standard}
\begin{algorithmic}
\Function{MI}{input vector $x_0$, target class $t$, iterations $\alpha$, patience $\beta$, minimum cost threshold $\gamma$, learning rate for gradient descent $\lambda$} 
\For{$i\leftarrow 1, \, \ldots, \, \alpha$}
\State $x_i \leftarrow x_{i-1} - \lambda \cdot \nabla c(x_{i-1})$
\If{$c(x_i) \geq \max(c(x_{i-1}), \, \ldots, \, c(x_{i-\beta}))$}
    \State \textbf{break}
\EndIf
\If{$c(x_i)\leq \gamma$}
    \State \textbf{break}
\EndIf
\EndFor
\State \textbf{return} $[\argmin_{x_i} c(x_i), \min_{x_i} c(x_i)]$
\EndFunction
\end{algorithmic}
\label{algo:standardMI}
\end{algorithm}
In speaker recognition, every speaker denotes their own class.
Hence, \MI can reveal representations of the data from every single individual in the training data set.
Particularly, this data can encode biometric as well as paralinguistic features.



\section{Sliding \MI Attack}
\label{sec:sliding_model_inversion}
In this section, we present our novel sliding \MI attack.
It extends standard \MI (see Algorithm~\ref{algo:standardMI}) to sequentially and chunk-wise processed data, e.g., audio data.
Instead of using \MI to invert every chunk of data separately, our sliding ML iteratively inverts overlapping chunks.
This way, some of the input to the \MI is already inverted, and hence, in wave-form similar to actual speech data.
Thereby, \MI can more successfully invert it into representatives of the original speech data. 

Our sliding \MI consists of the following steps: 
(1) We invert the first window of a randomly initialized input vector.
Note that different random initializations yield inversions of different quality. 
We experiment with several different types of initialization settings as our first main experiment, described in more detail in Section~\ref{ssec:exp1}.
(2) Then, we replace the first window of the input vector by the resulting inverted data.
(3) Next, we iteratively calculate the inverted data for the subsequent input window.
This input's first part consists of the previously inverted data, its second part stems from the random input vector.
Note that the amount of overlap for the sliding window determines the proportion of the previously inverted data and the randomly initialized input vector that are used for calculating the inversion during the \MI attack. 
Its value depends on the \emph{stride} of our inversion.
For our experiments, we use a stride of 500 samples (roughly 30ms).
Since in this new method, updates rely on the output of the previous inversion, we cannot use parallelization as a speed-up.
Instead, the stride determines the computational overhead. 
By increasing the stride value, computational time can be decreased.

\begin{figure}[t]
    \centering
    \includegraphics[trim=0 5.5cm 0 0,clip,width=\linewidth, page=2]{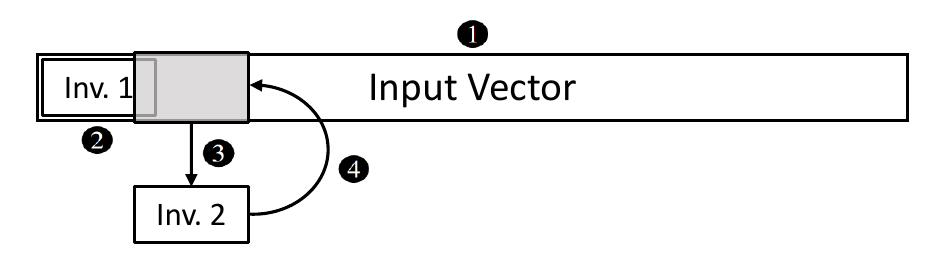}
    \caption{\textbf{Sliding \MI.}
    \encircle{1} We initialize a random input vector. \encircle{2} Starting from the beginning, we invert the first window based on this vector and replace the vector's first part by our inverted data.
    \encircle{3} For the all subsequent windows, we use parts of the previously inverted vector and fill the remainder with the input vector to apply \MI. \encircle{4} We then iteratively replace the input vector with our inverted data.
    }
    \label{fig:sliding_model_inversion}
\end{figure}

For a visualisation of our novel approach, see Figure~\ref{fig:sliding_model_inversion}.
Note that in addition to the hyperparameters of the standard MI, we need to specify the length of our input data $l$, the stride $s$, and the window size $w$.
While $s$ specifies the overlap between subsequent inversions, $w$ determines the lengths of the data chunks that are inverted.
For each chunk, inversion is performed as an iterative process as in the standard MI.
Since the beginning and the end of the inverted vector are iterated less, we cut the returned vector to half the window size.
See Algorithm~\ref{algo:slidingMI} for a formal introduction of our novel sliding ML.

\begin{algorithm}[tb]
\caption{\emph{Sliding} Model Inversion Attack}\label{alg:cap}
\begin{algorithmic}
\Function{SMI}{target class $t$, length $l$, stride $s$, windowsize~$w$, $\alpha, \beta, \gamma, \lambda$ as in Algorithm~\ref{algo:standardMI}}
\State inverted[0,\ldots,$l$] $\leftarrow [\mathcal{N}(\mu,\,\sigma^{2})]^l$
\For{$k\leftarrow 0, \, \ldots, \, (l-s) $ \textbf{according to stride} s}
\State $x \leftarrow \text{inverted}[k:(k+w)]$
\For{$i\leftarrow 1, \, \ldots, \, \alpha$}
\State $x_i \leftarrow x_{i-1} - \lambda \cdot \nabla c(x_{i-1})$
\If{$c(x_i) \geq \max(c(x_{i-1}), \, \ldots , \, c(x_{i-\beta}))$}
    \State \textbf{break}
\EndIf
\If{$c(x_i)\leq \gamma$}
    \State \textbf{break}
\EndIf
\EndFor
\State inverted$[k:(k+w)] \leftarrow \argmin_{x} c(x)$
\EndFor
\State \textbf{return} inverted$[\frac{w}{2} : (l-\frac{w}{2})]$
\EndFunction
\end{algorithmic}
\label{algo:slidingMI}
\end{algorithm}

\section{Experiments}
\label{sec:mi-sincnet}
We conduct three experiments:
The \emph{first experiment} is similar to \MI in other domains, i.e., it inverts random input vectors back to the original input data domain.
In the \emph{second experiment} we do not invert the whole NN, but only the layers up to the d-vectors, which provide unique voice features of an individual (see Figure~\ref{fig:experiments}). Inverting these vectors instead of full audio samples reduces the computational costs of the attack.
Our \emph{third experiment} shows that in speaker recognition, our \MI attack enables us to impersonate individual speakers, and to synthesize speech samples for them.
The spoofing is performed based on the inverted audio samples from the experiment one.

We attack the NN-based speaker recognition system using SincNet \cite{ravanelli2018speaker}, trained on the TIMIT dataset \cite{garofolo1993timit}, and we use the pretrained model provided by Ravanelli et al.~\cite{ravanelli2018speaker}.
To perform our \MI attack, we assume the attacker to have white-box access to the target model.
This is the case, for example, when the speaker recognition model is deployed to a user-device, e.g. for biometric identification.
Further, the attacker needs a unique identifier of their target individual under attack to know which class of the training data to invert.
This can be, for example, the name or some pseudonymized combination of characters, in the case of the TIMIT dataset, e.g. ``FGMB0''.


We quantify the success of our experiments as follows:
\begin{itemize}
    \item \emph{Percentage of correctly classified inverted samples}.
    We quantify the classification accuracy of the original target speaker model on both the inverted audio data and the inverted d-vectors. 
    An inverted sample is ``correctly classified'' if it is classified as the correct original speaker.
    \item \emph{Euclidean distance between original and reconstructed d-vector}.
    We measure the Euclidean distance between both d-vectors to specify the similarity between the respective samples. 
\end{itemize}
The first metric allows us to analyse if the \MI may be considered successful with respect to the target model, i.e., it answers the question of how successfully this model can be fooled.
The second metric, in contrast, focuses on the inverted samples' similarity to the original samples.
Hence, it quantifies the similarity from the perspective of a human listener.

Since \MI generates average representations of training classes, we use the target model's classification accuracy on averaged per-speaker samples as a baseline (97.84\% and 75.97\% of correct classification on train and test data, respectively). 
In the following, we present our overall experimental setup to then describe every experiment in more detail.

\subsection{Experiment 1: Invert Audio Samples}
\label{ssec:exp1}

In the first experiment, we use \MI to calculate full inverted audio samples which could be used to trick the speaker recognition model under attack without human listeners present.
The experiment is designed to answer the following three questions:
(1) Is it possible to successfully generate inverted audio samples for speaker recognition?
(2) Which kind of randomly initialized input vector to the \MI attack produces the most successful inverted audio samples (with respect to the classification as the original speaker)?
(3) How does our new sliding \MI approach improve the results in comparison to standard \MI?

\paragraph*{Experiment.}
Over all experiments, the audio data chunks are 3200 samples (or 200ms) long, and our sliding \MI uses a stride of 500 samples (roughly 30ms).
We evaluate the following (random) initializations:
\begin{itemize}
    \item \emph{Plain inputs}: all zeros, all ones, and all minus ones;
    \item \emph{Noises}: white, pink, brown, violet, and blue noise. 
    We generated them with methods pre-implemented in the \texttt{python-acoustics} library and applied the tanh-function to transform them to the interval range $\left[-1, 1\right]$ with zero mean in a non-linear manner;
    \item \emph{Samplings from distributions}, such as uniform (ranges $[0, 1]$ or $[-1, 1]$), Gaussian (with $\mu=0$ or $\sigma=0.2$), Laplace (with $\mu=0$ or $b=0.07$), Gumbel (with $\mu=0$ or $\beta=0.1$), and von Mises (with $\mu=0$ or $\kappa=0.1$);
    \item \emph{Samples from another dataset}: Librispeech \cite{panayotov2015librispeech}, used as a plain input or averaged over 50, 100, 150 input vectors or with white noise ($0.85 \cdot \text{input vector} + 0.15 \cdot \text{noise}$). 
\end{itemize}


For the optimization process within the \MI algorithm (see~Section~\ref{ssec:modelinv}), we optimize two parameters, namely the maximum number of iterations $\alpha$ and the learning rate $l$. 
For all experiments, setting $\alpha=1000$ showed to be sufficient.
The optimal $l$ depends on the experiment and is reported in the results.
In our evaluation, for every input with its settings and optimal $l$, we report the following metrics:
\begin{itemize}
    \item \emph{MI Accuracy}: the percentage of correctly classified inverted samples;
    \item \emph{\# Correct Speakers}:
    the number of correctly classified speakers;
    \item \emph{Avg. Eucl. Distance with Std. Deviation}: the average Euclidean distance of the inverted sample to original samples of the same speaker in the d-vector space, calculated on the successfully inverted audio samples.
\end{itemize}
For the Euclidean distances within the d-vector space, the average within-speaker distance in the original training samples can serve as a baseline ($1.923 \cdot 10^{-1}$).

\paragraph*{Results.}
\textit{(1) Is it possible to successfully generate inverted audio samples for speaker recognition?}
By looking at the distribution of the inverted samples, we observe that it does not fully match the distribution of the original data.
Since the inverted samples also sound differently from the original ones---they do not necessary sound like speech---hence, they do not allow to fool a \emph{human} listener. 
However, the results suggest that standard \MI is good enough to fool the classification of the \emph{automatic} speaker recognition system with an accuracy of up to $54.76\%$.
The classification accuracy can be significantly improved to $90.48\%$ through our new sliding \MI.
For speaker recognition models in charge of identity control for a highly secured system, this accuracy on inverted data would be beyond acceptable.
Our novel sliding \MI, also reduces the Euclidean distance between inverted and original samples for some input vectors, in comparison to a standard \MI.
However, despite this decrease in distance, the reconstructed speech samples still do not sound very close to the original speaker.

\textit{(2) Which kind of randomly initialized input vector to the MI attack produces the most successful inverted audio samples (with respect to the classification as the original speaker)?}
We can also conclude that not all input vectors to the \MI are equally suited to create inverted audio samples which successfully fool the speaker model:
plain input vectors achieve the lowest quality in inversion with respect to classification accuracy.
We assume that this is due to the difficulty of transforming constant vectors into speech-like wave forms through optimization.
It seems that random initialization or data that is already in wave form are more suited inputs to \MI for audio data: 
The best classification accuracy can be achieved with white noise and tanh activation. 
Brown noise exhibits the poorest performance, yet it exhibits a relatively small mean Euclidean distance.
The Laplace distributions achieves the overall highest results.
See Table~\ref{tab:comparisonStandardSliding} for an overview of results.

\textit{(3) How does our new sliding MI approach improve the results in comparison to standard MI?}
We observe from the results in Table~\ref{tab:comparisonStandardSliding} that sliding MI exhibits a higher performance than standard MI.
While with standard MI, the accuracy of the target model on the inverted data is $54$\%, depending on the random initialization, our sliding MI yields above $90$\% accuracy.

\subsection{Experiment 2: Partial \MI to Invert d-Vectors}
While in previous applications on other data types, only a complete \MI back to the original input domain is valuable, this is different for speaker recognition:
the d-vectors, which are feature representations of the voice samples, already carry important paralinguistic information that can, for example, be used to generate spoofed audio samples \cite{jia2018transfer}.
Inverting simple d-vectors instead of full audio samples reduces the computational costs of the attack (since it does not need to be performed sequentially), and can be performed with standard \MI attacks.
Therefore, in our second experiment we set out to invert a model on the intermediate layers with the aim to answer the following two questions: (1) Is it possible to successfully invert d-vector?
(2) Which input vector produces the most successful inverted d-vector (with respect to the classification as the original speaker)?

\paragraph*{Experiment.}
We apply partial inversion by removing the SincNet and MLP part of the network and only focusing on the submodel~3 for the d-vector inversion (see Figure~\ref{fig:experiments}). 

\begin{figure}[t]
\centering
\scalebox{0.25}{
    \input{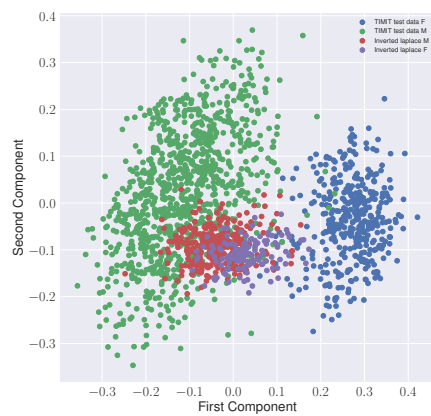} 
    }
    \caption{\textbf{PCA on d-Vectors.} PCA fitted on the TIMIT test dataset (\emph{blue}: female; \emph{green}: male) and used to transform the inverted d-vectors (\emph{purple}: female; \emph{red}: male). 
    Results indicate that inverted d-vectors reveals the individuals' gender.}
    \label{fig:pca_gender}
\end{figure}

\paragraph*{Results.}
\textit{(1) Is it possible to successfully invert d-vector?}
Our findings show that we can reconstruct d-vectors that are successfully classified as the original speaker (classification accuracy of the target model on them reaches $100\%$).
This outperforms the baseline where we measure accuracy of the target model on averaged per-speaker samples ($97.84\%$).
However, even for the best-performing input (zeros), the Euclidean distance of 0.84 $\pm$ 0.028 is clearly above the baseline average within speaker distance ($1.923 \cdot 10^{-1}$).
To evaluate whether the inverted d-vectors still leak the individual speakers' privacy, we perform a principal component analysis (PCA) and train a binary classifier to predict the individuals' gender.
We fit the PCA to the TIMIT test dataset and transform the inverted d-vectors. 
Our results are visualized in Figure~\ref{fig:pca_gender}.
They suggest that the inverted d-vector leak gender privacy.

\textit{(2) Which input vector produces the most successful inverted d-vector (with respect to the classification as the original speaker)?}
Since d-vectors and audio data have different properties, they require different input vectors for successful inversion. 
Sound noises are good inputs for audio samples.
However, our observations suggest that initializing the input vector with sound noises does not yield high-quality inverted d-vector representations.
Instead, we found plain zeros perform best when inverting d-vectors.

\begin{table*}[t]
    \centering
    \begin{tabular}{cccrcc}
    \toprule
    & \textbf{Sample Type} & \textbf{Learning R.} & \textbf{MI Accuracy} & \textbf{\# Correct Speakers} & \textbf{Avg. E. Dist. $\pm$ Std. Dev.} \\ 
    \midrule
    \multirow{4}{*}{\rotatebox[origin=c]{90}{\parbox[c]{1cm}{\centering plain inputs}}} 
    & \multirow{2}{*}{ones}       & {\color{gray}$1 \cdot 10^{-05}$} & {\color{gray}0.43\% }& {\color{gray}2 }& {\color{gray}0.800 $\pm$ 0.0416 }\\
    & & 0.2 & 5.41\% & 25 & 0.696 $\pm$ 0.0663 \\
    & \multirow{2}{*}{zeros}      & {\color{gray}$1 \cdot 10^{-08}$ }& {\color{gray}0.65\% }& {\color{gray}3 }& {\color{gray}0.792 $\pm$ 0.0377 }\\ 
    & & 0.5 & 8.87\% & 41 & 0.720 $\pm$ 0.0682 \\ \hline
    \multirow{4}{*}{\rotatebox[origin=c]{90}{\parbox[c]{1.4cm}{\centering dists}}} 
    & \multirow{2}{*}{Gumbel}     & {\color{gray}0.01 }& {\color{gray}54.76\% }& {\color{gray}253 }& {\color{gray}0.797 $\pm$ 0.0562 }\\
    &      & 0.01 & 89.83\% & 415 & 0.748 $\pm$ 0.0561 \\ 
    & \multirow{2}{*}{Laplace}    & {\color{gray}0.005 }& {\color{gray}54.33\% }& {\color{gray}251 }& {\color{gray}0.793 $\pm$ 0.0555 }\\
    & &0.005 & 90.48\% & 418 & 0.752 $\pm$ 0.0550 \\
    \hline 
    \multirow{4}{*}{\rotatebox[origin=c]{90}{\parbox[c]{1.4cm}{\centering noise}}}  
    & \multirow{2}{*}{white-tanh} & {\color{gray}0.2 }& {\color{gray}54.11\% }& {\color{gray}250 }& {\color{gray}0.798 $\pm$ 0.0549 }\\ 
    && 0.2 & 88.31\% & 408 & 0.757 $\pm$ 0.0561 \\
    & \multirow{2}{*}{brown}      & {\color{gray}0.01 }& {\color{gray}5.84\% }& {\color{gray}27 }& {\color{gray}0.676 $\pm$ 0.0938 }\\
    && 0.05 & 31.82\% & 147 & 0.681 $\pm$ 0.0679 \\ \hline 
    \multirow{8}{*}{\rotatebox[origin=c]{90}{\parbox[c]{1.4cm}{\centering Librispeech samples}}} 
    & \multirow{2}{*}{Librispeech sample} & {\color{gray}0.001 }& {\color{gray}6.277\% }& {\color{gray}29 }& {\color{gray}0.751 $\pm$ 0.0768 }\\
    & & 0.2 & 22.73\% &105 & 0.696 $\pm$ 0.0558 \\
    & \multirow{2}{*}{Sample noise + Librispeech sample} & {\color{gray}0.005 }& {\color{gray}52.6\% }& {\color{gray}243 }& {\color{gray}0.779 $\pm$ 0.0673 }\\
    & & 0.01 & 90.26\% & 417 & 0.757 $\pm$ 0.0595 \\
    & \multirow{2}{*}{Librispeech mean} & {\color{gray}0.001 }& {\color{gray}33.55\% }& {\color{gray}155 }& {\color{gray}0.750 $\pm$ 0.0665 }\\
    & & 0.01 & 56.06\% & 259 & 0.707 $\pm$ 0.0713 \\
    & \multirow{2}{*}{Sample noise + Libspeech mean} & {\color{gray}0.005} & {\color{gray}45.89\%} & {\color{gray}212} & {\color{gray}0.776 $\pm$ 0.0604 }\\ 
    & & 0.01 & 80.30\% & 371 & 0.757 $\pm$ 0.0600 \\ 
    \bottomrule
    \end{tabular}
    \caption{\textbf{Results for standard \MI ({\color{gray}gray}) and sliding \MI (black)} calculated for inverted audio samples of the 462 speakers in the TIMIT dataset (326 men, 136 women).
    Sliding \MI improves standard \MI (higher accuracy and lower average Euclidean distance.)}
    \label{tab:comparisonStandardSliding}
    \vspace{-6mm}
\end{table*}

\subsection{Experiment 3: Create Deepfakes}
\label{ssec:exp3}
Even though the inverted samples are classified correctly by the target model, they do not necessarily carry useful information for human listeners. 
With the following experiments, we focus on the question: 
(1) Based on the inverted audio samples, is it possible to generate audio data that resembles the original speaker for a human listener?
The experiment can be considered as a proof-of-concept to demonstrate further security risks in speaker recognition systems made possible by our attack.

\paragraph*{Experiment.}
To generate the deepfakes, we use the work from~\cite{jia2018transfer}.
Their architecture consist of three parts: 
(1) speaker encoder, (2) speech synthesizer, and (3) vocoder. 
The speaker encoder is used to create a d-vector out of an audio file, which characterizes the audio sample in vector space. 
With this information, the speech synthesizer creates the mel-spectogram by using the d-vectors. 
Finally, vocoder grabs the mel-spectogram to perform frequency to time domain conversion of. 
The underlying speech synthesizer is Tacotron 2 \cite{shen2018natural} and vocoder is Wavenet \cite{oord2016wavenet}. 
We use the inverted audio samples from our sliding \MI, and the inverted d-vectors as input for the method. 
In principle, inverted audio samples can be fed directly into the speaker encoder for the deepfake generation.
However, to use our inverted d-vectors (2048 dimensions), we have to transform them to match the deepfake model's speaker encoding, as it expects d-vectors with 256 dimensions. 
To do so, we train an MLP to map one vector space to another. 
The MLP has two hiddens layers  with 1024 and 512 neurons and an output layer with 256 neurons.
We use tanh activation in the first two layers.
We create the training set for this transformation by feeding sound files to our and the deepfake's speaker encoder.
Our encodings are used as input while their encodings are treated as the outputs to be learned.



\paragraph*{Results.}
\textit{(1) Based on the inverted audio samples, is it possible to generate audio data that resembles the original speaker for a human listener?}
As reported in experiment~2, our d-vectors, though correctly classified as the original speakers, were far away from the speaker's original d-vectors in the Euclidean space.
The question if a generated sample sounds similar to an original one is a semantic question and depends on the sensitivity of the context.
From the authors' perspective on inspection case-by-case, the d-vector-based deepfakes did not allow individual speakers characteristics to be recognized.
However, based on the inverted audio samples from our novel sliding \MI, we were able to generate a few good quality spoofed audio samples that resembled the original speaker.\footnote{Examples for the original audio data, averaged and inverted samples, and the spoofed audio data generated based on the inverted audio samples, are available at \url{https://www.dropbox.com/sh/ge6xx90laqmru9b/\ AABUsS3p4EwaN0n7g4Rgq8rwa}.}
With such samples at hand, an attacker could, hence, spoof someone's identity solely based the inverted data from the pre-trained NN.
We expect this to become even much more prevalent with more sophisticated deepfake generation systems in the future.

\section{Countermeasures and Discussion}
Speaker recognition systems heavily rely on learning individual per-speaker characteristics in order to fulfill the task they are designed for. 
Therefore, these systems always and necessarily contain information about the speaker data that they were trained on. 
Noising out individual speaker characteristics will result in drastically decreased performance of the systems.
In particular, pseudonomization~\cite{noe2022towards} and privacy methods that are used in general speech systems (e.g., \cite{aloufi2021tandem,jin2009voice,srivastava2020evaluating, srivastava2020design, qian2018hidebehind,srivastava2019privacy}) render speaker recognition unusable for their original purpose.


\paragraph*{Protecting against MI attacks.}
As an alternative, one can consider privacy protection methods that aim at impeding \MI attacks. 
Most existing defenses from other domains focus on suppressing the model confidence score or reducing their utility.
This can be done by injecting uniform noise to them~\cite{salem2020updates}, reducing their precision~\cite{fredrikson2015model} or their dispersion~\cite{yang2020defending}.
The latter one leads to a decrease in the correlation between the input data and the scores, which renders \MI attacks more inaccurate.
With a similar aim, the use of regularization in the training loss function has been reported as a defense~\cite{wang2020improving}. 
Additionally, hardware-oriented solutions to prevent an attacker from accessing the model parameters to decrease \MI success, or at least preventing the extraction of intermediate features (see our experiment~2) can be applied~\cite{xu2020midas}. 

\paragraph*{Differential privacy.}
Initial work empirically showed that Differential Privacy (DP)~\cite{dwork2006differential}
can reduce the success of \MI attack's~\cite{fredrikson2015model} when using a very large amount of noise, which, in return, drastically degrades the model's performance.
Later work suggests that DP training for ML models cannot at all prevent \MI attacks~\cite{zhang2020secret} because its aim is to dissimulate the presence of a data point in a specific data set and not to protect privacy over classes of data.


\paragraph*{Limitations.}
So far, \MI attacks require the availability of an NN's confidence scores, and the attack's success depends largely on the random initializations.
Especially, many speaker recognition tools depend on the cosine similarity \cite{zhang2022mfa,desplanques2020ecapa} for which the algorithm would need to be updated.
Also, the quality of the spoofed audio samples is limited by the deepfake creation methods.
As a consequence, the practical impact of our attack might currently still be limited.
However, with new and ever more powerful privacy attacks and deepfake methods being proposed, the threat space of exploiting privacy attacks to violate security of speaker recognition systems will gain importance.
It is, hence, important to create awareness and to consider and protect privacy and security jointly, rather than separately.

\section{Conclusion}
In this work, for the first time, we successfully perform MI attacks on audio data.
Therefore, we introduce a novel sliding MI method which leverages the sequential properties of the audio data for improved inversion.
We experimentally evaluate the attack's success on a state-of-the-art speaker recognition system.
Our results indicate that our inverted audio samples can be used as a departure point for further attacks against the security of the target system.
Thereby, we highlight the importance of implementing adequate privacy protection in such systems.

\section{Acknowledgements}
This research was supported by the Bavarian Ministry of Economic Affairs, Regional Development and Energy.
The authors acknowledge J.~Williams for comments on the manuscript.

\bibliographystyle{IEEEtran}
\bibliography{mybib}

\end{document}